\documentclass[12pt]{revtex4}
\usepackage{latexsym,color,amsmath,graphicx,amssymb,subfigure}
\renewcommand{\epsilon}{\varepsilon}
\newcommand{\bra}[1]{\mbox{$\langle #1 |$}}
\newcommand{\ket}[1]{\mbox{$| #1 \rangle$}}

\begin{document}
\title{Quantum key distribution with an unknown and untrusted source}
\author{Yi Zhao, Bing Qi, and Hoi-Kwong Lo}
\affiliation{Center for Quantum Information and Quantum Control,
Department of Physics and Department of Electrical \& Computer
Engineering, University of Toronto, Toronto, Ontario, M5S 3G4,
Canada.}
\begin{abstract}
The security of a standard bi-directional ``plug \& play'' quantum
key distribution (QKD) system has been an open question for a long
time. This is mainly because its source is equivalently controlled
by an eavesdropper, which means the source is unknown and untrusted.
Qualitative discussion on this subject has been made previously.  In
this paper, we present the first quantitative security analysis on a
general class of QKD protocols whose sources are unknown and
untrusted. The securities of standard BB84 protocol, weak+vacuum
decoy state protocol, and one-decoy decoy state protocol, with
unknown and untrusted sources are rigorously proved. We derive
rigorous lower bounds to the secure key generation rates of the
above three protocols. Our numerical simulation results show that
QKD with an untrusted source gives a key generation rate that is
close to that with a trusted source.
\end{abstract}
\maketitle

\section{Introduction}
Quantum key distribution (QKD) \cite{BB84,Gisin:RMP,Ekert91}, when
combined with the one-time pad algorithm, provides unconditional
communication security. The unconditional security is rigorously
proved based on fundamental physics principles such as quantum
no-cloning theorem and Heisenberg's uncertainty principle
\cite{SecurityProofs} rather than unproven computational complexity
assumptions. The unconditional security of QKD has been proven even
when implemented on imperfect practical set-ups with coherent laser
sources and semi-realistic models \cite{GLLP,ILM}.

Unconditional security of quantum cryptography is different from
``absolute security". ``Unconditional'' in the security proof of QKD
 means that we are not making any assumption about Eve's technology,
except that quantum mechanics is correct. However, we do have to
make assumptions on Alice's and Bob's sides to ensure the security.
The concept of unconditional security in QKD is discussed in details
in \cite{Review:Framework}.

Recently, the ideas of device-independent security proofs of QKD and
security from causality constraints have been proposed
\cite{Security:DeviceIndependent,Security:Causality,Security:NoSignalingQKD},
but a complete proof of unconditional security along those lines is
still missing. Moreover, any such device-independent security
proofs, even if successfully constructed in future, will not be
applicable practical QKD systems due to the well-known detection
efficiency loophole. This loophole can be filled under the fair
sampling assumption. Unfortunately, the fair sampling assumption can
be invalid in practical QKD set-ups due to some imperfections, like
the detection efficiency mismatch. Indeed, the detection efficiency
mismatch opens a back door for several practical attacks, including
the faked states attack
\cite{ThrHack:Makarov_Mismatch,ThrHack:Makarov_FakedStatesSARG04}
and the time-shift attack \cite{ThrHack:TimeShift}. The latter
attack has even been experimentally demonstrated on a commercial QKD
system \cite{ExpHack:TimeShift}, thus highlighting the weakness of
practical QKD systems.

It is very important to develop security proofs with testable
assumptions, and test the assumptions both theoretically and
experimentally. For example, the assumption of phase randomization
is often made in security proofs of practical set-ups. However, the
phases of signals are not naturally randomized in practice.
Fortunately, the validity of the phase-randomization assumption can
be confidently guaranteed by actively randomizing the phase of each
signal, which has only been demonstrated in a recent experiment
\cite{Zhao:APL07}. See, however, \cite{Lo:NonRandomPhase} for a
security proof that does not require the phase randomization
assumption.

The validity of the coherent state assumption is also questionable.
For example, it is common to use pulsed laser diodes as sources in
QKD experiments. These laser diodes are driven by pulsed electrical
currents. When the driving current is switched on, it will take a
short while before the laser's gain reaches its stabilizing
threshold. During this transition period, the output from the diode
cannot be viewed as coherent state. Therefore, it is not rigorous to
consider the entire pulse as a coherent state.

A more severe problem comes from the standard bi-directional
(so-called ``plug \& play'') design \cite{Stucki:NJP2002}, which is
widely used in commercial QKD systems. In this particular scheme,
bright pulses are generated by Bob (a receiver) rather than Alice (a
sender). The pulses will travel through the channel, which is fully
controlled by Eve (an eavesdropper), before entering Alice's lab to
get encoded and sent back to Bob. Eve can perform arbitrary
operation on the pulses when they are sent from Bob to Alice. In the
worst case, Eve can replace the original pulses by her own
sophisticatedly prepared optical signals. Such an attack is called
the Trojan horse attack \cite{Gisin:TrojanHorse}. Therefore, it is
highly risky to assume that Alice uses a coherent state source in
the security analysis of ``plug \& play'' QKD systems.

Previously, a qualitative argument on the security of bi-directional
QKD system was provided in \cite{Gisin:TrojanHorse}. The intuition
is to show that by applying heavy attenuation, an input state with
arbitrary photon number distribution can be transformed into an
output state with Poisson-like distribution. However, it is
challenging to quantify how close to the Poissonian state the output
state is.

We start from another intuition: we look into the actual photon
number distribution created by the internal loss of Alice's local
lab. The phase randomization can transform arbitrary input state
into a classical mixture of number states \cite{Gisin:TrojanHorse}.
By modeling the internal loss inside Alice's local lab as a beam
splitter, for each particular input photon number, the photon number
of output state obeys binomial distribution. Note that this is not a
binomial-like, but a rigorous binomial distribution. The analysis of
binomial distribution is in general harder than that of Poisson
distribution. However, in this way we can quantitatively and
rigorously analyze its security.

The discovery of decoy methods can dramatically improve the
performance (by means of higher key rate and longer transmission
distance) of coherent laser based QKD systems
\cite{Decoy:Hwang,Decoy:LoPRL,Decoy:LoISIT,Decoy:Practical,Decoy:WangPRL,Decoy:WangPRA,Decoy:Harrington,Decoy:Twoway}.
The decoy method has been experimentally demonstrated over long
distances
\cite{Decoy:ZhaoPRL,Decoy:ZhaoISIT,Decoy:144km,Decoy:PanPRL,Decoy:LATES,Decoy:60Hour,Decoy:YuanAPL,Decoy:130km}.

In decoy state QKD, each bit is randomly assigned as a signal state
or one of the decoy states. Each state has its unique average photon
number. These states can be prepared by setting different internal
transmittances $\lambda$ in Alice's local lab. For example, if a bit
is assigned as a signal state, the internal transmittance for this
bit will be $\lambda_S$. If a bit is assigned as a decoy state, the
internal transmittance for this bit will be
$\lambda_D\neq\lambda_S$. Normally $\lambda_D<\lambda_S$.

In previous analysis on decoy state QKD
\cite{Decoy:LoPRL,Decoy:LoISIT,Decoy:Practical,Decoy:Twoway}, one
important assumption is that the yield of $n$ photon state $Y_n$ in
signal state is the same as $Y_n$ in decoy state. i.e.,
$Y_n^S=Y_n^D$. Here $Y_n$ is defined as the conditional probability
that Bob's detectors generate a click given that Alice sends out an
$n$ photon signal. This is true because in the analysis of
\cite{Decoy:LoPRL,Decoy:LoISIT,Decoy:Practical,Decoy:Twoway} Eve
knows only the output photon number $n$ of each pulse. Another
fundamental assumption is that the quantum bit error rate (QBER) of
$n$ photon state $e_n$ in signal state is the same as $e_n$ in decoy
state. i.e., $e_n^S=e_n^D$. Note that, once Eve knows some
additional information about the source, the above two fundamental
assumptions will \emph{fail} \cite{Decoy:WangInexactSource}.

We emphasize that in the case of ``Plug \& Play'' QKD, Eve knows
both the input photon number $m$ and the output photon number $n$.
Therefore she can perform an attack that depends on the values of
both $m$ and $n$. In Section \ref{subse:W+V} and Appendix
\ref{app:yn}, we show explicitly that $Y_n^S \neq Y_n^D$ and $e_n^S
\neq e_n^D$ in this case. The parameters that are the same for both
the signal state and the decoy states are $Y_{m,n}$ (the conditional
probability that Bob's detectors click given that this bit enters
Alice's lab with photon number $m$ and emits from Alice's lab with
photon number $n$) and $e_{m,n}$ (the QBER of bits with $m$ input
photons and $n$ output photons).

In brief, there is more information available to Eve once she
controls the source. The security analysis for decoy state QKD in
this case is much more challenging.


In this paper, we analyze the most general case: we consider the
source as controlled by Eve. Therefore the source is completely
unknown and untrusted. Rather surprisingly, we show that even in
this most general case, the security of the QKD system can be
analyzed quantitatively and rigorously. We also show that the decoy
method can still be used to enhance the performance of the system
dramatically when the source is unknown and untrusted. For the first
time, we show quantitatively that the security of ``plug \& play''
QKD system is understandable and achievable. Moreover, we show what
measures are necessary to ensure the security of the QKD system, and
rigorously derive a lower bound of the secure key generation rate.
Our numerical simulation results show that QKD with an untrusted
source gives a key generation rate that is close to that with a
trusted source.

It is important to implement QKD with testable assumptions. In this
paper, we showed that the coherent source assumption can be removed.
Nonetheless, we still keep a few standard assumptions including
single mode assumption, phase randomization assumption, etc. in our
security proof. To ensure that our assumptions of single-mode and
phase randomization are satisfied in practice, we propose specific
experimental measures for Alice to implement. More concretely, we
propose that Alice uses a strong filter to filter out other optical
modes and uses active phase randomization to achieve phase
randomization. It would be interesting to see the security
consequence of removing, say, the single mode assumption. However,
this is beyond the scope of this work.

This paper is organized in the following way: in Section II, we
propose some measures that should be included in the QKD set-up, and
a key term -- ``untagged bit'' -- is defined; in Section III, we
study the experimental properties of the untagged bits; in Section
IV, the photon number distribution for untagged bits is analyzed; in
Section V, we prove the security of practical QKD system with
unknown and untrusted source, and explicitly show the equation for
the key generation rate; in Section VI, we prove the security of two
decoy state protocols -- the weak+vacuum protocol and the one-decoy
protocol -- with unknown and untrusted sources; in Section VII,
numerical simulation results are shown; in Section VIII, we present
our conclusion and discuss future directions.

\section{Measures to enhance the security}

\begin{figure}
  \includegraphics[width=16cm]{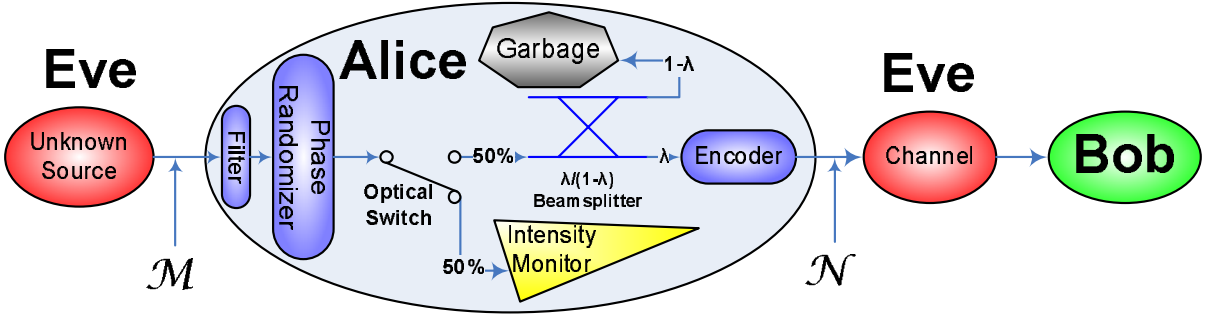}\\
  \caption{A schematic diagram of the set-up that coped with the
  three measures as suggested: Filter is used to guarantee the single mode
  assumption; Phase Randomizer is used to guarantee the phase
  randomization assumption; Optical Switch and
  Intensity Monitor
  are used to randomly sample the photon number of input pulses.
    All the internal losses inside Alice's
  local lab is modeled as a $\lambda/(1-\lambda)$ beam splitter. That is,
  any input photon has $\lambda$ probability to get encoded and sent from Alice
  to Bob, and $1-\lambda$ probability to be discarded into the
  Garbage. $\mathcal{M}$ and $\mathcal{N}$ are the random variables
  for input photon numbers and output photon numbers,
  respectively. Note that in a standard ``plug \& play'' setup, the actual source
  is inside Bob's local lab. However, Eve can replace the pulses
  sent by Bob with arbitrary optical signals. This is equivalent to the
  general case in which Eve controls the source.}\label{FIG:setup}
\end{figure}

Here we will use three measures, which were briefly mentioned in
\cite{Gisin:TrojanHorse}, to enhance the security of the system. A
general system that has applied these measures is shown in Fig.
\ref{FIG:setup}. There are various sources of losses inside Alice's
apparatus. Here we model all the losses as a $\lambda/(1-\lambda)$
beam splitter. That is, the internal transmittance of Alice's local
lab is $\lambda$. We assume that Alice can set $\lambda$ accurately
via, say variable optical attenuator. In other words, for any photon
that enters the encoding arm, it has a probability $\lambda$ to get
encoded and sent out from Alice.

\begin{enumerate}
\item
We pointed out and demonstrated in \cite{ExpHack:TimeShift} that the
side-channel can be exploited by Eve to acquire additional
information. To shut down these side-channels, we need to place a
filter (Filter in Fig. \ref{FIG:setup}) which works in spectral,
spatial, and temporal domains. In other words, only pulses of the
desired mode can pass through the filter. Therefore, we can use
single mode assumption for each signal. Incidentally, the single
mode assumption may not hold for an open-air QKD set-up. This is
because 1) the free space will not suppress the propagation of
higher modes and 2) the collection system at Bob's side can only
collect part of the beam sent from Alice.

\item The phase randomization is a general assumption made in most
security proofs on practical set-ups \cite{ILM,GLLP,Decoy:LoPRL}. It
can disentangle the input pulse from Eve by transforming it into a
classical mixture of Fock states $\sum_{n=0}^\infty
p_n\ket{n}\bra{n}$ \cite{Gisin:TrojanHorse}. Its feasibility has
been experimentally demonstrated \cite{Zhao:APL07}. Alice should
apply the phase randomization on the input optical signals. In Fig.
\ref{FIG:setup}, this is accomplished by the Phase Randomizer.

\item We need to monitor the pulse energy to acquire some information
about the photon number distribution. By randomly sampling a portion
of the pulses to test the photon numbers, we can estimate some
bounds on the output photon number distribution as shown in the
following sections. In Fig. \ref{FIG:setup}, this is accomplished by
the Optical Switch and the Intensity Monitor.
\end{enumerate}


Suppose that $2K$ pulses entered Alice's local lab, within which $K$
pulses were randomly chosen by the Optical Switch in Fig.
\ref{FIG:setup} for testing photon numbers (these pulses are called
``sampling bits''), and the rest $K$ pulses were encoded and sent to
Bob (these pulses are called ``coding bits''). Define the pulses
with photon number $m\in[(1-\delta)N,(1+\delta)N]$ as ``untagged''
bits, and pulses with photon number $m<(1-\delta)N$ or
$m>(1+\delta)N$ as ``tagged'' bits. Note that the definitions of
``untagged'' and ``tagged'' here are different from those in
\cite{GLLP}. From random sampling theorem (see, like, \cite{QIQC})
we know that the probability that there are less than $K\Delta$
tagged sampling bits and more than $(\Delta+\epsilon)K$ tagged
coding bits is asymptotically less than $e^{-O(\epsilon^2K)}$.
$\epsilon$ should be chosen under the condition that $\epsilon^2
K\gg1$. Therefore there are no less than $(1-\Delta-\epsilon)K$
untagged coding bits with high fidelity.

In the following discussion, we will focus on these
$(1-\Delta-\epsilon)K$ untagged bits. Of course, there can also be
some untagged bits in the rest $(\Delta+\epsilon)K$ bits, but
neglecting these out-of-scope untagged bits just makes our analysis
conservative.

 $N$ and $\delta$ can, in principle, be arbitrarily chosen.
However, some constraints will be applied to optimize the key
generation rate. We will discuss the optimal choice later.

\section{Properties of the untagged bits}

In QKD experiments, the two most important measurable outputs are
the gain \cite{Gain} and the QBER. In our analysis, we are more
interested in the gain and the QBER of the untagged bits. This is
because the input photon numbers of the untagged bits are
concentrated within a narrow range, making it much easier to analyze
the security.

However, Alice cannot in practice perform quantum non-demolishing
(QND) measurement on the photon number of the input pulses with
current technology. Therefore, she does not know which bits are
tagged and which are untagged. As a result, the gain \cite{Gain} $Q$
and the QBER $E$ of the untagged bits cannot be measured
experimentally. Here $Q$ is defined as the \emph{conditional}
probability that Bob's detector clicks given that Alice sends out an
untagged bit and Alice and Bob use the same basis; $E$ is defined as
the \emph{conditional} probability that Bob's bit value is different
from Alice's given that Bob's detector clicks, Alice sends out an
untagged bit, and Alice and Bob use the same basis.

In an experiment, Alice and Bob can measure the overall gain $Q_e$
and the overall QBER $E_e$. The subscript $e$ denotes the
experimentally measurable overall properties. Moreover, they know
the probability that certain bit to be tagged or untagged from the
above analysis. Although they cannot measure the gain $Q$ and the
QBER $E$ of the untagged bits directly, they can estimate the upper
bounds and lower bounds of them. The upper bound and lower bound of
$Q$ are
\begin{equation}\label{eq:Qbound}
\begin{aligned}
    \overline{Q} &= \frac{Q_e}{1-\Delta-\epsilon},\\
    \underline{Q} &= \max(0,
    \frac{Q_e-\Delta-\epsilon}{1-\Delta-\epsilon}).
\end{aligned}
\end{equation}
The upper bound and lower bound of $E\cdot Q$ can be estimated as
\begin{equation}\label{eq:Ebound}
    \begin{aligned}
    \overline{E\cdot Q} &=
    \frac{Q_e E_e}{1-\Delta-\epsilon},\\
    \underline{E\cdot Q} &=
    \max(0,\frac{Q_e E_e-\Delta-\epsilon}{1-\Delta-\epsilon}).
    \end{aligned}
\end{equation}

To get tighter bounds on $Q$ and $E\cdot Q$, we need to minimize
$\Delta$, which means that $\delta$ should be made large so as to
minimize the amount of tagged bits. See, however, discussion after
Eqs. \eqref{eq:pnbound}.

\section{Photon number distribution of untagged bits}

Consider an untagged bit with input photon number $m\in[(1-\delta)
N, (1+\delta) N]$. The conditional probability that $n$ photons are
emitted by Alice given that $m$ photons enter Alice obeys binomial
distribution as
\begin{align}\label{eq:pn}
    P_n(m) &= {m\choose n} \lambda^n(1-\lambda)^{m-n}. & (0\le\lambda\le1)
\end{align}


For untagged bits (i.e., $m\in[(1-\delta) N, (1+\delta) N]$), we can
show that the upper bound and lower bound of $P_n(m)$ are:
\begin{equation}\label{eq:pnbound}
\begin{aligned}
    \overline{P_n} &= \left\{
                        \begin{array}{ll}
                         (1-\lambda)^{(1-\delta)N}, & \hbox{if
                         $n=0$;}\\
                          {(1+\delta)N\choose n}\lambda^n(1-\lambda)^{(1+\delta)N-n}, & \hbox{if $1\le n \le (1+\delta)N$;} \\
                          0, & \hbox{if $n>(1+\delta)N$;}
                        \end{array}
                      \right.\\
    \underline{P_n} &= \left\{
                        \begin{array}{ll}
                        (1-\lambda)^{(1+\delta)N}, & \hbox{if
                        $n=0$;}\\
                          {(1-\delta)N\choose n}\lambda^n(1-\lambda)^{(1-\delta)N-n}, & \hbox{if $1\le n \le (1-\delta)N$;} \\
                          0, & \hbox{if $n>(1-\delta)N$;}
                        \end{array}
                      \right.
\end{aligned}
\end{equation}
under \textbf{Condition 1:}
\begin{equation}\label{eq:condition1}
(1+\delta)N\lambda<1.
\end{equation}

Condition 1 suggests that the expected output photon number of any
untagged bit should be lower than 1. This is easy to implement
experimentally. For example, for $N=10^6$, Alice can simply set
$\lambda=10^{-7}$ so that the expected output photon number is
$0.1$. Most reported BB84 implementations satisfy Condition 1.

To get tighter bounds on $P_n(m)$, we need to minimize $\delta$.
However, as we discussed below Eq. \eqref{eq:Ebound}, minimizing
$\delta$ will lower the amount of untagged bits (i.e., there will be
fewer pulses contain photon number $m\in[(1-\delta)N,(1+\delta)N]$
as the bound becomes narrower), thus loosening the bounds on the
gains and QBERs of untagged bits. As a summary, there is a trade-off
between the tightness of the bounds of $P_n(m)$ and the tightness of
the bounds of $Q$ and $E\cdot Q$. The optimal choice of $\delta$
depends on the properties of specific system, and can be obtained
numerically.


%

\section{Generalized GLLP Results with Untrusted Source}
From the work of Gottesman-Lo-L\"{u}tkenhaus-Preskill (GLLP)
\cite{GLLP}, the secure key generation rate of standard BB84
protocol \cite{BB84} is given by
\begin{equation}\label{eq:rateGLLP}
R \ge
\frac{1}{2}\{-Q_ef(E_e)H_2(E_e)+\underline{Q\Omega}[1-H_2(\frac{Q_eE_e}{\underline{Q\Omega}})]\},
\end{equation}
where $1/2$ is the probability that Alice and Bob use the same
basis, $Q_e$ and $E_e$ are obtained experimentally, $f(>1)$ is the
bi-directional error correction inefficiency
\cite{Security:ErrorCorrectionEfficiency}, and
\begin{equation}\label{eq:omega}
\Omega=1-\frac{P_M}{Q},
\end{equation}
where $P_M=\sum_{n=2}^{\infty} P_n(m)$ is the probability of output
multiphoton signals.
 Recall that if
the input photon number $m=(1+\delta)N$, we have
$$
P_n((1+\delta)N) = \left\{
  \begin{array}{ll}
    \underline{P_n}, & \hbox{if $n=0$;} \\
    \overline{P_n}, & \hbox{if $n\ge1$.}
  \end{array}
\right.
$$
Therefore $\underline{P_0}+\sum_{n=1}^\infty \overline{P_n}=1$. The
upper bound of $P_M$ is $\overline{P_M}=\sum_{n=2}^{\infty}
\overline{P_n}=1-\underline{P_0}-\overline{P_1}$, and the lower
bound of $\Omega$ is
$$
\underline{\Omega}=1-\frac{\overline{P_M}}{\underline{Q}}.
$$
The lower bound of $Q\Omega$ is thus given by
\begin{equation}\label{eq:QOmegaBound}
    \underline{Q\Omega}=\underline{Q}-\overline{P_M}=\underline{Q}+\underline{P_0}+\overline{P_1}-1,
\end{equation}
where $\underline{Q}$ can be obtained via Eq. \eqref{eq:Qbound}.

Plugging Eq. \eqref{eq:QOmegaBound} into Eq. \eqref{eq:rateGLLP}, we
have the key generation rate per bit sent by Alice, given an
untrusted source is used, as
\begin{equation}\label{eq:rateGLLP_Generalized}
R \ge
\frac{1}{2}\{-Q_ef(E_e)H_2(E_e)+(\underline{Q}+\underline{P_0}+\overline{P_1}-1)[1-H_2(\frac{Q_eE_e}{\underline{Q}+\underline{P_0}+\overline{P_1}-1})]\}.
\end{equation}

The numerical simulation of the above analysis is presented in
Section \ref{se:simulation}.

\section{Combining with decoy states}
Decoy method
\cite{Decoy:Hwang,Decoy:LoPRL,Decoy:LoISIT,Decoy:Practical,Decoy:WangPRL,Decoy:WangPRA,Decoy:Harrington}
significantly improves the performance for QKD systems with coherent
state source. Here, we will show that the idea of decoy states can
also be useful when the source is unknown and untrusted.

\subsection{Weak+vacuum protocol}\label{subse:W+V}
Among all the decoy state protocols, the weak+vacuum protocol is the
most popular one. It is shown to be the optimal protocol in
asymptotic case \cite{Decoy:Practical}. ``Asymptotic'' here means
infinitely long source data sequence. The weak+vacuum protocol has
been used in most experimental decoy state QKD implementations
\cite{Decoy:ZhaoISIT,Decoy:PanPRL,Decoy:144km,Decoy:LATES,Decoy:60Hour}.

In weak+vacuum protocol, there are three states: the signal state
(for which the internal transmittance of Alice is $\lambda_S$), the
weak decoy state (for which the internal transmittance of Alice is
$\lambda_D<\lambda_S$), and the vacuum state (for which the internal
transmittance of Alice is 0). We consider that only the signal state
is used to generate the final key, while the decoy states are solely
used to test the channel properties.

The error correction will consume
\begin{equation}\label{eq:ErrorCorrection}
    r_\text{EC}=Q_{e}^Sf(E_{e}^S)H_2(E_{e}^S)
\end{equation}
bit per signal sent from Alice, where $Q_{e}^S$ and $E_{e}^S$ are
the overall gain and overall QBER of signal state, $H_2$ is binary
Shannon function.

The probability that Alice sends out an untagged signal which is
securely transmitted to Bob is
\begin{equation}\label{eq:PrivacyAmplification}
    r_\text{PA} = (1-\Delta-\epsilon)Q_1^S[1-H_2(e_1^S)],
\end{equation}
where $Q_1^S$ and $e_1^S$ are the gain and the QBER of single photon
state in untagged bits. This is because Alice and Bob can, in
principle, measure the input photon number $m$ and the output photon
number $n$ accurately and therefore post-select the untagged bits
with $n=1$. They can then use these post-selected single-photon
untagged bits to generate the secure key. In practice, QND
measurements on $m$ and $n$ by Alice are not feasible with current
technology. However, Alice and Bob know the probability of certain
bit to be untagged. They can use random-hashing method to perform
privacy amplification to distill the secure key. Similar technique
was used in \cite{GLLP}.

The key generation rate in standard BB84 protocol is therefore given
by
\begin{equation}\label{eq:rateWV}
R \ge \frac{1}{2}(r_{PA}-r_{EC}) \ge
\frac{1}{2}\{-Q_{e}^Sf(E_{e}^S)H_2(E_{e}^S)+(1-\Delta-\epsilon)\underline{Q_1^S}[1-H_2(\overline{e_1^S})]\}
,
\end{equation}
where $1/2$ is the probability that Alice and Bob use the same
basis.

$Q_{e}^S$, $E_{e}^S$, $\Delta$, and $\epsilon$ can be determined
experimentally. Our main task is to estimate $\underline{Q_1^S}$ and
$\overline{e_1^S}$.


In previous analysis on decoy state QKD
\cite{Decoy:LoPRL,Decoy:LoISIT,Decoy:Practical,Decoy:Twoway}, one
important assumption is that the yield of $n$ photon state $Y_n$ in
signal state is the same as $Y_n$ in decoy state. i.e.,
$Y_n^S=Y_n^D$. Here $Y_n$ is defined as the conditional probability
that Bob's detectors generate a click given that Alice sends out an
$n$ photon signal. This is true because in the analysis of
\cite{Decoy:LoPRL,Decoy:LoISIT,Decoy:Practical,Decoy:Twoway} Eve
knows only the output photon number $n$ of each pulse. However, as
we will show below, this assumption is no longer valid in the case
that the source is controlled by Eve.

The key point is that Eve knows both the input photon number $m$ and
the output photon number $n$ when she controls both the source and
the channel. Therefore she can perform an attack that depends on the
values of both $m$ and $n$. In this case, the parameter that is the
same for these states is $Y_{m,n}$, the conditional probability that
Bob's detectors click given that this bit enters Alice's lab with
photon number $m$ and is emitted from Alice's lab with photon number
$n$. In this case, $Y_n$ is given by (see Appendix \ref{app:yn} for
details)
\begin{equation}\label{eq:Yn}
    Y_n = \sum_m P\{m|n\}Y_{m,n},
\end{equation}
where $P\{m|n\}$ is the conditional probability that the signal
enters Alice's local lab with photon number $m$ given that it is
emitted from Alice's lab with photon number $n$. Note that
$P\{m|n\}$ is dependent on the internal transmittance of Alice's
apparatus $\lambda$. Since $\lambda_S\neq\lambda_D$, we know that
$Y_n^S\neq Y_n^D$.

Another fundamental assumption for previous decoy state security
studies \cite{Decoy:LoISIT,Decoy:LoPRL,Decoy:Practical} is that the
QBER of $n$-photon state $e_n$ is the same for signal state and
decoy state. i.e., $e_n^S=e_n^D$. Unfortunately, from a similar
analysis as above, we can show that $e_n^S \neq e_n^D$ if Eve
controls the source. The parameter that is the same for the signal
state and the decoy states is $e_{m,n}$.

As a brief summary, in decoy state QKD, if the source is in Alice's
local lab and is solely accessible to Alice (that is, the source is
\emph{trusted}), we have $Y_n^S=Y_n^D$ and $e_n^S=e_n^D$, whereas if
the source is out of Alice's local lab and is accessible to Eve
(that is, the source is \emph{untrusted}), we have
$Y_{m,n}^S=Y_{m,n}^D$ and $e_{m,n}^S=e_{m,n}^D$.

The dependence of $Y_n$ and $e_n$ on different states (signal state
or one of the decoy states) is a fundamental difference between
decoy state QKD with untrusted source and decoy state QKD with
trusted source. In the latter case, the independence of $Y_n$ and
$e_n$ on different states  is a very powerful constraint on Eve's
ability of eavesdropping. However, this constraint is removed once
the source is given to Eve.

Eve's control over the source removes the two fundamental
assumptions in \cite{Decoy:LoISIT,Decoy:LoPRL,Decoy:Practical}. Eve
is given significantly greater power, and the security analysis is
much more challenging. However, rather surprisingly, it is still
possible to achieve the
 unconditional security quantitatively even if the source is given to
Eve. This is mainly because we are only focusing on the untagged
bits, whose input photon numbers are concentrated in a relatively
narrow range. Therefore we are still able to estimate
$\underline{Q_1^S}$ and $\overline{e_1^S}$.


\textbf{Proposition 1:} the lower bound of $Q_1^S$ for untagged bits
is given by
\begin{equation}\label{eq:q1boundWV}
    \begin{aligned}
    Q_1^S > \underline{Q_1^S}= \underline{P_1^S}\frac{\underline{Q^D}\underline{P_2^S}-\overline{Q^S}\overline{P_2^D}+(\underline{P_0^S}\overline{P_2^D}-\overline{P_0^D}\underline{P_2^S})\overline{Q^V}-\frac{2\delta
   N(1-\lambda_D)^{2\delta
   N-1}\underline{P_2^S}}{[(1-\delta)N+1]!}}{\overline{P_1^D}\underline{P_2^S}-\underline{P_1^S}\overline{P_2^D}}
    \end{aligned}
\end{equation}
under \textbf{Condition 2:}
\begin{equation}
\frac{\lambda_S}{\lambda_D}>\frac{(1+\delta)N-2}{(1-\delta)N-2}\left[\frac{(1+\delta)N-2}{2\delta
N}\right]^\frac{2\delta
N}{(1-\delta)N-2}\left[\frac{(1+\delta)N-2}{(1-\delta)N-2}\cdot\frac{e^2}{2\delta
N}\right]^\frac{1}{2[(1-\delta)N-2]}.
\end{equation}
Here $Q^S$, $Q^D$ and $Q^V$ are the gains of untagged bits of the
signal state, the decoy state, and the vacuum state, respectfully.
Their bounds can be estimated from Eqs. \eqref{eq:Qbound}. The
bounds of the probabilities can be estimated from Eqs.
\eqref{eq:pnbound}. Note that Condition 2 is easy to meet. For
example, in the numerical simulation in Section \ref{se:simulation},
we chose $N=10^6$ and $\delta=0.01$. In this case we can calculate
Condition 2 as $\frac{\lambda_S}{\lambda_D}>1.104$, which is very
reasonable to meet experimentally. Actually, $\lambda_S/\lambda_D$
is usually greater than 2 in previous decoy state QKD
implementations
\cite{Decoy:ZhaoPRL,Decoy:ZhaoISIT,Decoy:144km,Decoy:PanPRL,Decoy:LATES,Decoy:60Hour,Decoy:YuanAPL,Decoy:130km}.

\textbf{Proof:} Appendix \ref{app:q1}.



\textbf{Proposition 2:} the upper bound of $e_1^S$ for untagged bits
 is given by
\begin{equation}\label{eq:e1boundWV}
    e_1^S\le\overline{e_1^S}=\frac{\overline{E^SQ^S}-\underline{P_0^S}\underline{E^VQ^V}}{\underline{Q_1^S}},
\end{equation}
in which $E^S$ and $E^V$ are the QBERs of untagged bits of the
signal and the vacuum states, respectively. $\overline{E^SQ^S}$ and
$\underline{E^VQ^V}$ can be estimated from Eqs. \eqref{eq:Ebound}.
$\underline{P_0^S}$ can be estimated  by Eqs. \eqref{eq:pnbound}.
$\underline{Q_1^S}$ is given by Eq. \eqref{eq:q1boundWV}.

\textbf{Proof:} Appendix \ref{app:e1}.

Plugging Eqs. \eqref{eq:q1boundWV} \& \eqref{eq:e1boundWV} into Eq.
\eqref{eq:rateWV}, we can easily calculate the overall key
generation rate of weak+vacuum decoy state QKD protocol given the
source is under Eve's control.
%

%

\subsection{One-decoy protocol (asymptotic case)}
The one-decoy protocol is the simplest decoy state protocol. In the
one-decoy protocol, there are only two states: a signal state and a
weak decoy state. It can be viewed as a simplified version of the
weak+vacuum protocol since it does not have the vacuum state.

The one-decoy protocol is of practical interest, particularly due to
the difficulty of preparing perfect vacuum state. It has also been
widely used in experiments
\cite{Decoy:ZhaoPRL,Decoy:YuanAPL,Decoy:130km}.

Here, we will show that the one-decoy protocol is also applicable
when the source is under Eve's control in the asymptotic case. The
asymptotic case means that Alice sends infinitely long bit sequence
($K\sim\infty$).

In the one-decoy protocol, there is no vacuum state. Therefore we
cannot measure $Q_e^V$ or $E_e^V$, which means we cannot use Eqs.
\eqref{eq:Qbound} to estimate $\overline{Q^V}$ in Eq.
\eqref{eq:q1boundWV} or use Eqs. \eqref{eq:Ebound} to estimate
$\overline{E^VQ^V}$ in Eq. \eqref{eq:e1boundWV}. Nonetheless, we can
still estimate $\underline{Q_1^S}$ and $\overline{e_1^S}$.

\textbf{Proposition 3:} In absence of the vacuum state, a lower
bound of $Q_1^S$ and an upper bound of $e_1^S$ for untagged bits are
given by
\begin{equation}\label{eq:q1e1_OneDecoy}
    \begin{aligned}
    \underline{Q_1^S} &= \underline{P_1^S}\frac{\underline{Q^D}\underline{P_2^S}-\overline{Q^S}\overline{P_2^D}+(\underline{P_0^S}\overline{P_2^D}-\overline{P_0^D}\underline{P_2^S})\frac{\overline{E^SQ^S}}{\underline{P_0^S}E^V}-\frac{2\delta
   N(1-\lambda_D)^{2\delta
   N-1}\underline{P_2^S}}{[(1-\delta)N+1]!}}{\overline{P_1^D}\underline{P_2^S}-\underline{P_1^S}\overline{P_2^D}},\\
    \overline{e_1^S} &= \frac{\overline{E^S\cdot
    Q^S}}{\underline{Q_1^S}},
    \end{aligned}
\end{equation}
respectively under Condition 2 in the asymptotic case. Here $Q^S$
and $Q^D$ are the gains of untagged bits of the signal state and the
decoy state, respectively. Their bounds can be estimated from Eqs.
\eqref{eq:Qbound}. $E^S$ is the QBER of untagged bits of the signal
state. $\overline{E^S\cdot Q^S}$ can be estimated from Eqs.
\eqref{eq:Ebound}. $E^V=0.5$ in the asymptotic case. The bounds of
the probabilities can be estimated from Eqs. \eqref{eq:pnbound}.

\textbf{Proof:} Appendix \ref{app:one-decoy}.

Plugging Eqs. \eqref{eq:q1e1_OneDecoy} into Eq. \eqref{eq:rateWV},
we can easily calculate the overall key generation rate of one-decoy
protocol given the source is under Eve's control.

\section{Numerical Simulation with Coherent Source: Asymptotic
Case}\label{se:simulation}

 In the asymptotic case, Alice sends
infinitely long bit sequence ($K\sim\infty$). Therefore we can
consider $\epsilon\sim0$.
\subsection{Calculating $\Delta$}
For any $\delta\in[0,1]$, we can calculate $\Delta$ by
\begin{equation}\label{eq:deltap}
\Delta = 1-[\Phi(N+\delta N)-\Phi(N-\delta N)],
\end{equation}
where $\Phi$ is the cumulative distribution function of the photon
number for the input pulses.

Most QKD set-ups are based on coherent sources, which means that the
input photon number $m$ obeys Poisson distribution. It is natural to
set $N$ to be the average input photon number. For a Poisson
distribution centered at $N$, its cumulative distribution function
is given by
$$
    \Phi_p(x) = \frac{\Gamma(\lfloor x+1\rfloor,N)}{\lfloor
x\rfloor!},
$$
where $\Gamma(x,y)$ is the upper incomplete gamma function
$$
\Gamma(x,y) = \int_y^\infty t^{x-1}e^{-t}dt.
$$
It is complicated to calculate $\Phi_p(x)$ numerically, particularly
for large $x$. Therefore in numerical simulation, we approximate the
Poisson distribution by a Gaussian distribution centered at $N$ with
a variance $\sigma^2=N$. Note this is an excellent approximation for
large $N$. The Gaussian cumulative distribution function is given by
\begin{equation}\label{eq:cumulative_gaussian}
    \Phi_g(x) = \frac{1}{2}[1+\text{erf}(\frac{x-N}{\sqrt{2N}})],
\end{equation}
where
$$
\text{erf}(x) = \frac{2}{\sqrt{\pi}}\int_0^xe^{-t^2}dt
$$ is the error function. Notice that erf$(x)$ is an odd function,
from Eqs. \eqref{eq:deltap}\eqref{eq:cumulative_gaussian} we have
\begin{equation}\label{eq:deltap_gaussian}
    \begin{aligned}
        \Delta &= 1-[\Phi_g(N+\delta N)-\Phi_g(N-\delta N)]\\
                 &= 1-\frac{1}{2}[\text{erf}(\frac{\delta N}{\sqrt{2N}})-\text{erf}(\frac{-\delta
N}{\sqrt{2N}})]=1-\text{erf}(\sqrt{\frac{N}{2}}\delta).
    \end{aligned}
\end{equation}

\subsection{Simulating experimental outputs}

If the photon number of an input pulse obeys Poisson distribution
with average photon number $N$, the photon number of the output
signal also follows Poisson distribution with average photon number
$N\lambda$.

For a QKD setup with channel transmittance $\eta(=e^{-\alpha l}$,
where $\alpha$ is the loss coefficient and $l$ is the distance
between Alice and Bob), Bob's quantum efficiency $\eta_\text{Bob}$,
detector intrinsic error rate $e_\text{det}$ and background rate
$Y_0$, the gain and the QBER of the signals are expected to be
\cite{Decoy:Practical}
\begin{equation}\label{eq:qmuemu}
    \begin{aligned}
    Q_e &= Y_0+1-\exp(-\eta\eta_\text{Bob}N\lambda),\\
    E_e &=
\frac{e_0Y_0+e_\text{det}[1-\exp(-\eta\eta_\text{Bob}N\lambda)]}{Q_e}.
    \end{aligned}
\end{equation}
The experimental outputs are clearly determined by Alice's internal
transmittance $\lambda$ which needs to be set before the experiment.
In our simulation, the optimal values for $\lambda_S$ and
$\lambda_D$ are selected numerically via exhaustive search.

With these simulated experimental outputs, we can calculate the
lower bound of key generation rate from Eqs \eqref{eq:Qbound},
\eqref{eq:Ebound}, \eqref{eq:pnbound},
\eqref{eq:rateGLLP_Generalized}, \eqref{eq:rateWV},
\eqref{eq:q1boundWV}--\eqref{eq:qmuemu}.

\begin{figure}
  \includegraphics[width=12cm]{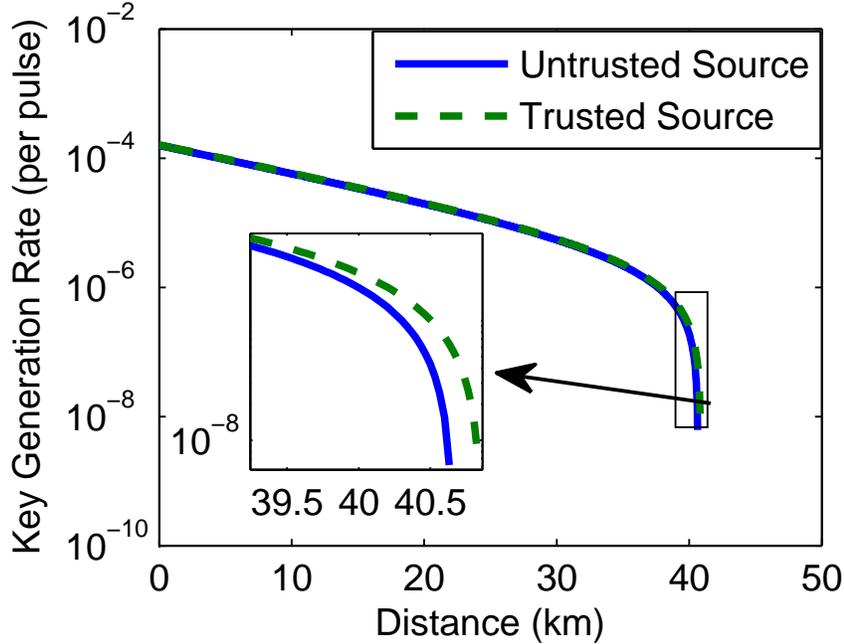}\\
  \caption{Simulation of GLLP protocol for $N=10^6$ and
  $\delta=1\%$. Citing GYS \cite{GYS} data. Inset: the magnified
  tail.
   The two cases (with a trusted source and with an untrusted
  source) give very similar results. We need to magnify the tail (see the
  inset) to see the slight advantage gained by using a trusted
  source. We calculated the ratio of the key
  generation rate with an untrusted source over that with a trusted
  source. The ratios are 98.0\%, 97.7\%, and 75.3\% at 0 km, 20 km,
  and 40km, respectively.}\label{FIG:RvsD_GLLP}
\end{figure}

\subsection{Simulation results}

Our simulation is based on the parameters reported by \cite{GYS} as
shown in Table \ref{Tab:simulation_parameter}.

\begin{table}[!b]
  \centering
  \caption{Simulation Parameter from GYS \cite{GYS}.}\label{Tab:simulation_parameter}
    \begin{tabular}{c c c c}
      \hline
      $\eta_\text{Bob}$ & $\alpha$ & $Y_0$ & $e_\text{det}$ \\
      4.5\% & 0.21dB/km & $1.7\times 10^{-6}$ & 3.3\% \\
      \hline
    \end{tabular}
\end{table}

We choose to set $N=10^6$, which is very reasonable: if the
wavelength is 1550 nm and the pulse repetition rate is 1 MHz, the
average input laser power will be $\sim0.128\mu$W, or $-38.9$ dBm.
Even if the channel loss from the source to Alice is 40dB
($\sim200$km telecom fiber), the required average output power from
the source is $\sim1.28$mW, which can be easily provided by many
commercial pulsed laser diodes. We chose $\delta$ to be 10 standard
deviations as $\delta=0.01$.

The simulation result for GLLP protocol is shown in Fig.
\ref{FIG:RvsD_GLLP}. We can see that the key generation rate with an
untrusted source is very close to that with a trusted source. Their
difference is almost negligible, and is only visible by magnifying
the tail (see the inset).

\begin{figure}
  \includegraphics[width=12cm]{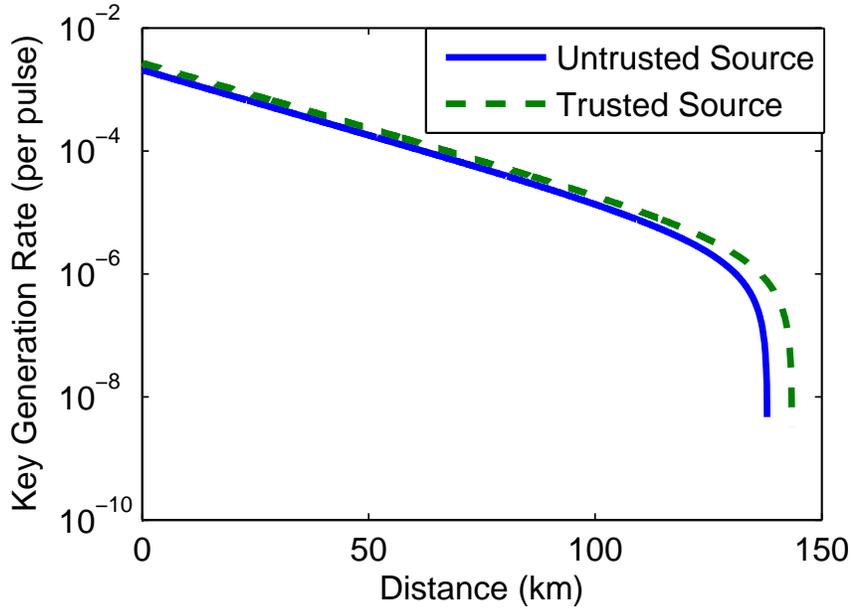}\\
  \caption{Simulation of weak+vacuum decoy state protocol for $N=10^6$ and
  $\delta=1\%$. Citing GYS \cite{GYS} data. The two cases (with a trusted source and with an untrusted
  source) give very close results. We calculated the ratio of the key
  generation rate with an untrusted source over that with a trusted
  source. The ratios are 77.3\%, 76.8\%, and 73.6\% at 0 km, 50 km,
  and 100km, respectively.
  }\label{FIG:RvsD_WV}
\end{figure}

The simulation result for weak+vacuum protocol is shown in Fig.
\ref{FIG:RvsD_WV}. We can see that the key generation rate with an
untrusted source is still very close to that with a trusted source.
By simply comparing the maximum transmission distances, we can see
that the difference is merely 5 km for weak+vacuum decoy state
protocol.

\begin{figure}
  \includegraphics[width=12cm]{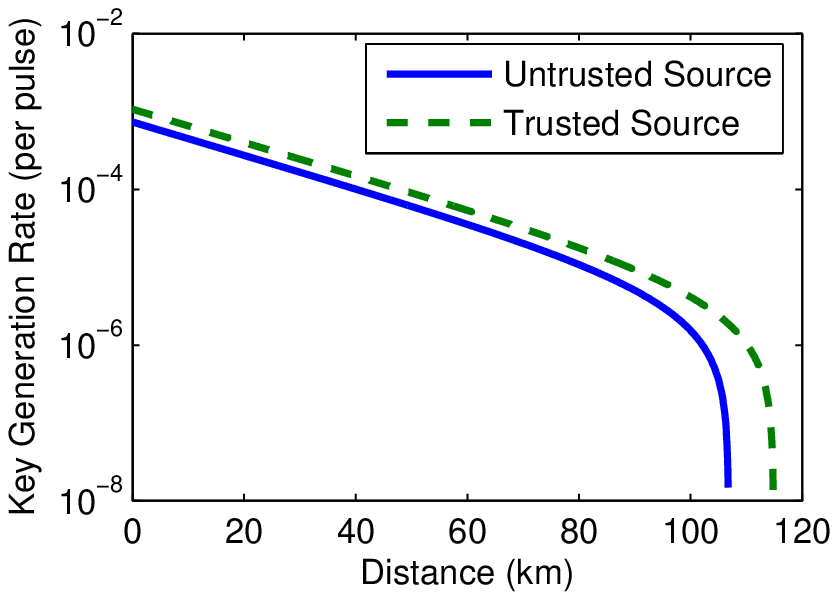}\\
  \caption{Simulation of one-decoy protocol for $N=10^6$ and
  $\delta=1\%$. Citing GYS \cite{GYS} data. The two cases (with a trusted source and with an untrusted
  source) give very close results. We calculated the ratio of the key
  generation rate with an untrusted source over that with a trusted
  source. The ratios are 68.6\%, 67.1\%, and 37.1\% at 0 km, 50 km,
  and 100km, respectively.}\label{FIG:RvsD_OneDecoy}
\end{figure}

The simulation result for one-decoy protocol is shown in Fig.
\ref{FIG:RvsD_OneDecoy}. We can see that the key generation rate
with an untrusted source is still very close to that with a trusted
source. The difference of maximum transmission distances is merely 8
km.

The above results are surprisingly good because we did not assume
any \emph{a priori} knowledge about the source in the security
analysis. In other words, Alice and Bob do not know the fact that
the source is Poissonian and therefore they cannot assume any photon
number distribution.

One important reason for achieving this high performance is that we
applied heavy attenuation on the input pulses. Note that the input
pulse has $\sim10^6$ photons, while the output pulse has less than
one photon on average. The internal attenuation of Alice's local lab
is greater than -60dB. We know that heavy attenuation will transform
arbitrary photon number distribution into a Poisson-like
distribution.

\begin{figure}
\subfigure[Simulation for $N=10^6$ and $\delta=0.4\%$.]{
\label{FIG:RvsD_Delta_0.004}
\includegraphics[width=8cm]{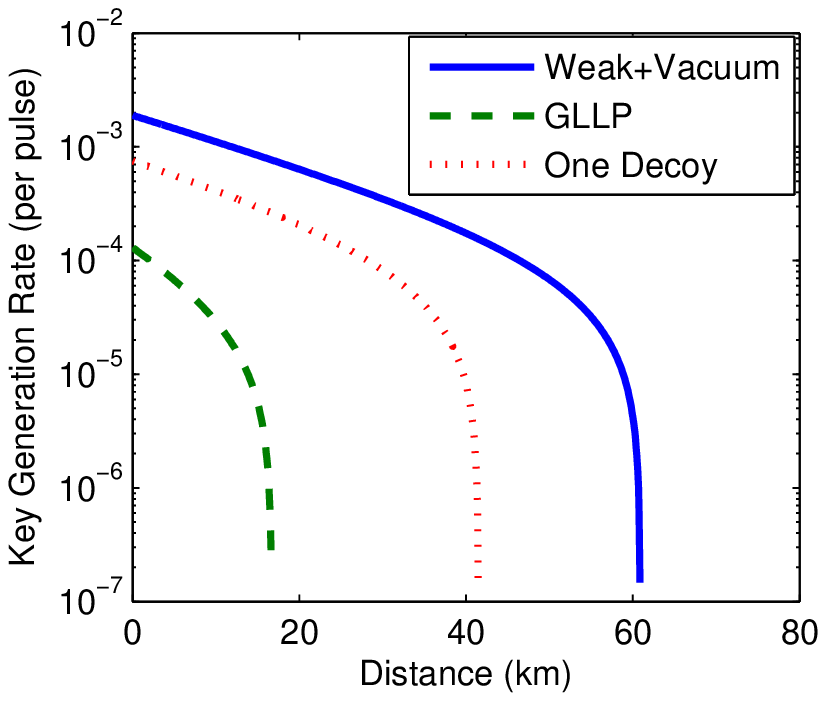}}
\subfigure[Simulation for $N=10^6$ and $\delta=11\%$.]{
\label{FIG:RvsD_Delta_0.2}
\includegraphics[width=7.8cm]{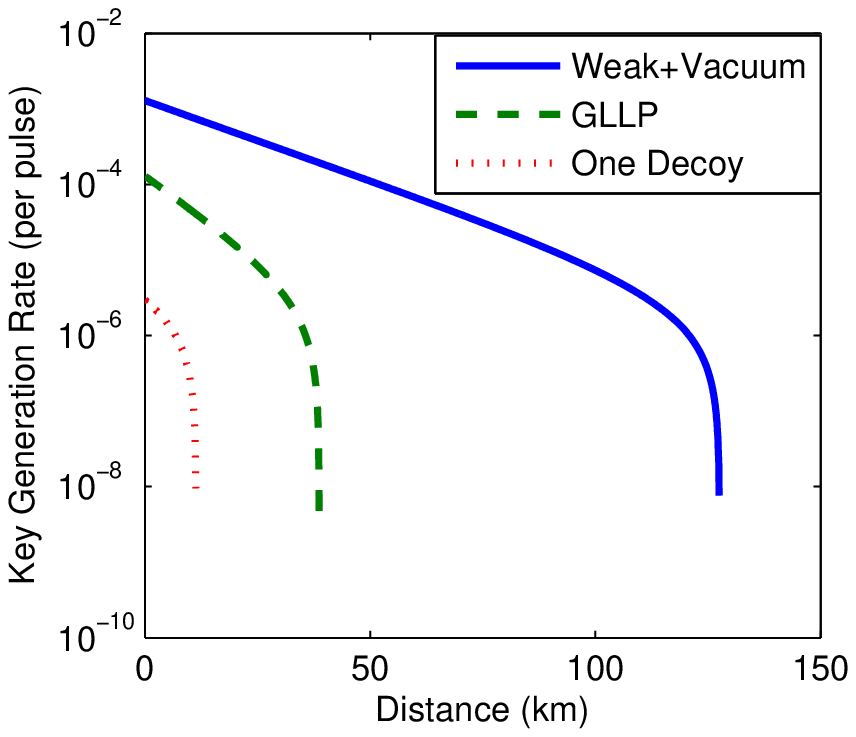}}\\
  \caption{Simulation results for too small (a) and too large (b) values of $\delta$.
  Citing GYS \cite{GYS} data. Both figures show key rates of different protocols with
  untrusted source.}\label{FIG:LargeDelta}
\end{figure}

As we mentioned before, $\delta$ can be arbitrarily chosen. However,
choosing $\delta$ too large or too small will make the security
analysis less optimal (i.e., conservative). Examples are given in
Fig. \ref{FIG:LargeDelta}. We can clearly see that inappropriate
choice of $\delta$ can deteriorate the performance of the system.
The one-decoy protocol with untrusted source is particularly
sensitive to the value of $\delta$.

\begin{figure}
  \includegraphics[width=12cm]{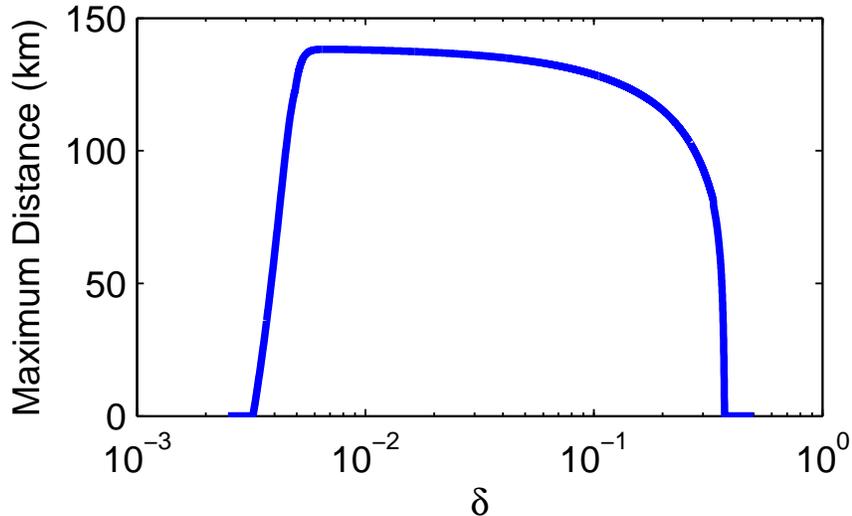}\\
  \caption{Maximum transmission distances of weak+vacuum protocol
   for various choices of $\delta$s.}\label{FIG:DvsDelta}
\end{figure}

The analytical optimization of $\delta$ can be complicated. Here we
just study this problem numerically. We calculated the maximum
possible transmission distances for different $\delta$. The results
are shown in Fig. \ref{FIG:DvsDelta}. We can clearly see that there
is an optimal choice of $\delta$. Note that our analysis is valid
for arbitrary value of $\delta$. The optimal value of $\delta$ will
give us the optimal (while still being rigorous) estimate on the
security of the system. Alice and Bob do not need to choose a
certain value of $\delta$ before the experiment. They only need to
find an optimal value of $\delta$ during the data post-processing.

The flat top in the curve of Fig. \ref{FIG:DvsDelta} suggests the
insensitivity of the maximum transmission distance on $\delta$ in a
wide range. We can see that the maximum transmission distance
changes only $8\%$ within the range of $\delta$ from 5 standard
deviations to 100 standard deviations. Therefore in practice, one
can simply set $\delta$ to be a few standard deviations and achieve
near-optimal results.

\section{Conclusion}
In this paper, we present the first rigorous quantitative security
analysis of a QKD system with an unknown and untrusted source. This
analysis is particularly important for the security of a standard
``Plug \& Play'' system. We showed that, rather surprisingly, even
with an unknown and untrusted source, unconditional security of QKD
system is still achievable, with and without the decoy method.
Moreover, we explicitly give the experimental measures that have to
be taken to ensure the security, and the theoretical analysis that
can be directly applied to calculate the final secure key generation
rate. One can easily extend our analysis to understand the security
of QKD network, in which the source is often untrusted.

For the first time, the unconditional security of the ``plug \&
play'' QKD system with current technology is made possible by us.
The ``plug \& play'' structure has clear advantage over
uni-directional structure since it does not require any active
compensation on the phase or the polarization. The self-compensating
property of the ``plug \& play'' structure makes it much simpler to
implement than the uni-directional structure, and makes it much
quieter (i.e., much lower QBER). All the commercial QKD systems
\cite{IDQ,MagiQ} are based on this simple and reliable structure.
However, the lack of rigorous security analysis has been an obstacle
for its development for a long time. With our straightforward
theoretical and experimental solution, we expect the ``plug \&
play'' structure to receive much more attention.

The security of practical QKD systems is a serious issue. Recently,
several quantum hacking works have been reported
\cite{ExpHack:TimeShift,ExpHack:DetectionTime}. It is very important
to implement QKD system based on tested assumptions. There are still
several crucial imperfections that are not analyzed in this paper.
For example, how can we understand the imperfection due to
non-single-mode (note that this is particularly important for
free-space QKD)? How can we analyze the fluctuation of internal
transmittance $\lambda$? Also, how can we test the key assumptions
in our analysis: $Y_{m,n}^S=Y_{m,n}^D$ and $e_{m,n}^S=e_{m,n}^D$?
These question marks suggest us a simple fact: although we are
approaching the unconditional security of practical QKD set-up, we
are not there yet.

We thank enlightening discussions with C. H. Bennett, C.-H. F. Fung,
D. Gottesman, D. F. V. James, X. Ma, J.-W. Pan, L. Qian, and X.-B.
Wang. Support of the funding agencies CFI, CIPI, the CRC program,
CIFAR, MITACS, NSERC, OIT, and PREA is gratefully acknowledged.

\bibliography{Z:/reference}

\begin{thebibliography}{42}
\bibitem[{Gai()}]{Gain}
\bibinfo{note}{The gain is defined to be the ratio of the number of receiver
  Bob's detection events to the number of signals emitted by sender Alice in
  the cases where Alice and Bob use the same basis. It depends mainly on the
  intensity of signal, channel transmittance, and Bob's quantum efficiency.}
\end{thebibliography}

\bibliographystyle{apsrev}

\begin{appendix}

\section{Derivation of $Y_n$}\label{app:yn}
We set $\mathcal{M}$ as the random variable of the input photon
number, $\mathcal{N}$ as the random variable of the output photon
number, and $\mathcal{C}$ as the random variable of Bob's detector
status (y = detection, n = no detection). $Y_n$ is then given by the
conditional probability
\begin{equation}\label{eq:Yn_app}
    Y_n = \text{Pr}\{\mathcal{C}=\text{y}|\mathcal{N}=n\},
\end{equation}
and $Y_{m,n}$ is given by the conditional probability
\begin{equation}\label{eq:Ymn_app}
    Y_{m,n} = \text{Pr}\{\mathcal{C}=\text{y}|\mathcal{N}=n \&
    \mathcal{M}=m\}.
\end{equation}
$Y_n$ can be expended as
\begin{align*}
    Y_n &= \text{Pr}\{\mathcal{C}=\text{y}|\mathcal{N}=n\}\\
        &=
        \frac{\text{Pr}\{\mathcal{C}=\text{y}\&\mathcal{N}=n\}}{\text{Pr}\{\mathcal{N}=n\}}\\
        &=
        \sum_{m=0}^\infty\frac{\text{Pr}\{\mathcal{C}=\text{y}\&\mathcal{N}=n\&\mathcal{M}=m\}}{\text{Pr}\{\mathcal{N}=n\}}\\
        &=\sum_{m=0}^\infty\frac{\text{Pr}\{\mathcal{N}=n\&\mathcal{M}=m\}}{\text{Pr}\{\mathcal{N}=n\}}\frac{\text{Pr}\{\mathcal{C}=\text{y}\&\mathcal{N}=n\&\mathcal{M}=m\}}{\text{Pr}\{\mathcal{N}=n\&\mathcal{M}=m\}}\\
        &=\sum_{m=0}^\infty\text{Pr}\{\mathcal{M}=m|\mathcal{N}=n\}\text{Pr}\{\mathcal{C}=\text{y}|\mathcal{N}=n\&\mathcal{M}=m\}\\
        &=\sum_{m=0}^\infty P\{m|n\}Y_{m,n}. &\Box
\end{align*}

\section{Estimate of $Q_1^S$}\label{app:q1}
From definition \cite{Gain}, we know that the gain of untagged bits
is given by
$$
Q 
= \sum_{m=(1-\delta)N}^{(1+\delta)N}\sum_{n=0}^\infty
P_\text{in}(m)P_n(m)Y_{m,n},
$$
where $P_\text{in}(m)$ is the probability that the input signal
contains $m$ photons (i.e., the ratio of the number of signals with
$m$ input photons over $K$), $P_n (m)$ is the conditional
probability that the output signal contains $n$ photons given the
input signal contains $m$ photons, and is given by Eq.
\eqref{eq:pn}.

 The gains for signal, decoy,
and vacuum states in untagged bits are therefore given by
\begin{equation}\label{eq:Q_WVdecoy}
\begin{aligned}
    Q^S &= \sum_{m=(1-\delta)N}^{(1+\delta)N}\sum_{n=0}^\infty P_\text{in}(m) P_n^S(m) Y_{m,n},\\
    Q^D &= \sum_{m=(1-\delta)N}^{(1+\delta)N}\sum_{n=0}^\infty P_\text{in}(m) P_n^D(m) Y_{m,n},\\
    Q^V &= \sum_{m=(1-\delta)N}^{(1+\delta)N} P_\text{in}(m) Y_{m,0},
\end{aligned}
\end{equation}
respectively. Here $P_n^{S/D}(m)$ is $P_n(m)$ for the signal/decoy
state. Their bounds can be estimated from Eqs. \eqref{eq:pnbound}.
$Q^{S/D/V}$ cannot be measured experimentally, but their upper
bounds and lower bounds can be estimated from Eqs.
\eqref{eq:Qbound}. Note that $\Delta^{S/D/V}$ should be determined
experimentally. In asymptotic case, $\Delta^S=\Delta^D=\Delta^V$. If
the bit sequence sent by Alice is finite, $\Delta^{S/D/V}$ may not
be exactly the same due to statistical fluctuation.



We know that
\begin{equation}\label{eq:q1z1}
Q_1^S=\sum_{m=(1-\delta)N}^{(1+\delta)N}P_\text{in}(m)P_1^S(m)Y_{m,1}\ge\underline{P_1^S}\sum_{m=(1-\delta)N}^{(1+\delta)N}P_\text{in}(m)Y_{m,1}=\underline{P_1^S}Z_1,
\end{equation}
in which $\underline{P_1^S}$ can be calculated from Eqs.
\eqref{eq:pnbound}, and $Z_1$ is defined as
\begin{equation}\label{eq:Z1}
    Z_1 = \sum_{m=(1-\delta)N}^{(1+\delta)N}P_\text{in}(m)Y_{m,1}.
\end{equation}
If we can put a lower bound on $Z_1$, we will be able to estimate
the lower bound of $Q_1^S$.

$Z_1$ clearly arises from the contribution of single photon signals.
A natural strategy is to find an appropriate linear combination of
$Q^S$ and $Q^D$, in which the multi-photon signal contribution is
minimized (while keeping it positive) so that we can set a lower
bound on it as zero. Among all the multi-photon signals, the two
photon signal has much greater weight than signals with more
photons. Therefore, we will try to eliminate the two-photon signal
contribution first. Note that we can easily estimate the
contribution of vacuum signals from $Q^V$ and $E^V$.

Eqs. \eqref{eq:pnbound} show that $\underline{P_n^S}\le P_n^S(m) \le
\overline{P_n^S}$ and $\underline{P_n^D}\le P_n^D(m) \le
\overline{P_n^D}$ for untagged bits. Combining them with Eqs.
\eqref{eq:Q_WVdecoy}, we have
\begin{equation}\label{eq:qs-qd}
\begin{aligned}
    Q^S\overline{P_2^D}-Q^D\underline{P_2^S} &=
    \sum_{m=(1-\delta)N}^{(1+\delta)N}P_\text{in}(m)\sum_{n=0}^\infty[P_n^S(m)\overline{P_2^D}-P_n^D(m)\underline{P_2^S}]Y_{m,n}\\
    &\ge\sum_{m=(1-\delta)N}^{(1+\delta)N}P_\text{in}(m)\sum_{n=0}^\infty[\underline{P_n^S}\overline{P_2^D}-\overline{P_n^D}\underline{P_2^S}]Y_{m,n}\\
    &=
    \sum_{n=0}^\infty[\underline{P_n^S}\overline{P_2^D}-\overline{P_n^D}\underline{P_2^S}]\sum_{m=(1-\delta)N}^{(1+\delta)N}P_\text{in}(m)Y_{m,n}\\
    &=
    a_0\sum_{m=(1-\delta)N}^{(1+\delta)N}P_\text{in}(m)Y_{m,0}+a_1\sum_{m=(1-\delta)N}^{(1+\delta)N}P_\text{in}(m)Y_{m,1}\\
    &+(\underline{P_2^S}\overline{P_2^D}-\overline{P_2^D}\underline{P_2^S})\sum_{m=(1-\delta)N}^{(1+\delta)N}P_\text{in}(m)Y_{m,2}+\sum_{n=3}^{(1-\delta)N}a_2(n)\sum_{m=(1-\delta)N}^{(1+\delta)N}P_\text{in}(m)Y_{m,n}+a_3\\
    &=
    a_0\sum_{m=(1-\delta)N}^{(1+\delta)N}P_\text{in}(m)Y_{m,0}+a_1\sum_{m=(1-\delta)N}^{(1+\delta)N}P_\text{in}(m)Y_{m,1}\\
    &+\sum_{n=3}^{(1-\delta)N}a_2(n)\sum_{m=(1-\delta)N}^{(1+\delta)N}P_\text{in}(m)Y_{m,n}+a_3\\
    &= a_0Z_0+a_1Z_1+\sum_{n=3}^{(1-\delta)N}a_2(n)Z_2(n)+a_3
\end{aligned}
\end{equation}
where
\begin{eqnarray}\label{eq:a0}
    a_0 &=&
    \underline{P_0^S}\overline{P_2^D}-\overline{P_0^D}\underline{P_2^S};\\\label{eq:a1}
    a_1 &=&
    \underline{P_1^S}\overline{P_2^D}-\overline{P_1^D}\underline{P_2^S};\\\label{eq:a2}
    a_2(n) &=&
    \underline{P_n^S}\overline{P_2^D}-\overline{P_n^D}\underline{P_2^S};\\\label{eq:a3}
    a_3
    &=&-\sum_{n=(1-\delta)N+1}^{(1+\delta)N}\overline{P_n^D}\underline{P_2^S}\sum_{m=(1-\delta)N}^{(1+\delta)N}P_\text{in}(m)Y_{m,n}=-\sum_{n=(1-\delta)N+1}^{(1+\delta)N}\overline{P_n^D}\underline{P_2^S}Z_3(n);
    \end{eqnarray}
and
\begin{eqnarray}
  Z_0 &=& \sum_{m=(1-\delta)N}^{(1+\delta)N}P_\text{in}(m)Y_{m,0} =
  Q^V;\\\label{eq:Z2}
  Z_2(n) &=& \sum_{m=(1-\delta)N}^{(1+\delta)N}P_\text{in}(m)Y_{m,n}; \\
  Z_3(n) &=& \sum_{m=(1-\delta)N}^{(1+\delta)N}P_\text{in}(m)Y_{m,n}.
\end{eqnarray}

 Note that in Eq. \eqref{eq:qs-qd}, when $n=2$, the term
 $\underline{P_n^S}\overline{P_2^D}-\overline{P_n^D}\underline{P_2^S}=0$,
 which means we have removed the contribution from the two-photon
 signals.
 Our strategy is clear now: the contribution of vacuum signals ($a_0Z_0$) can
 be easily bounded as $a_0$ can be calculated from Eqs. \eqref{eq:a0}
 and an upper bound of $Z_0$ is given by
 $\overline{Z_0}=\overline{Q^V}$, which can be calculated from Eqs. \eqref{eq:Qbound};
 the contribution of single photon signals ($a_1Z_1$) is
 to be estimated while we know the exact value of $a_1$; we need to put some bounds
  on the higher order terms
($a_2$ and $a_3$) to complete an estimate of $Z_1$. As we will show
below, $a_1$ is negative under certain condition. Therefore we
should put a lower bound on the
higher order terms to find the lower bound of $Z_1$. 

\textbf{Lemma 1}

$a_1$ is negative under \textbf{Condition 2a:}
$$\frac{\lambda_S}{\lambda_D}>\frac{(1+\delta)N-1}{(1-\delta)N-1}.$$

\textbf{Proof:}

Expand Eq. \eqref{eq:a1} we have
\begin{equation}\label{eq:a1bound}
    \begin{aligned}
        a_1 &=
    \underline{P_1^S}\overline{P_2^D}-\overline{P_1^D}\underline{P_2^S}\\
    &=
    (1-\delta)N\lambda_S(1-\lambda_S)^{(1-\delta)N-1}\frac{(1+\delta)N[(1+\delta)N-1]}{2}\lambda_D^2(1-\lambda_D)^{(1+\delta)N-2}\\
    &- (1+\delta)N\lambda_D(1-\lambda_D)^{(1+\delta)N-1}\frac{(1-\delta)N[(1-\delta)N-1]}{2}\lambda_S^2(1-\lambda_S)^{(1-\delta)N-2}\\
    &=
    N^2(1-\delta^2)\lambda_S^2\lambda_D^2(1-\lambda_S)^{(1-\delta)N-2}(1-\lambda_D)^{(1+\delta)N-2}\left[\frac{(1+\delta)N-1}{2\lambda_S}-\frac{(1-\delta)N-1}{2\lambda_D}-\delta
    N\right].
    \end{aligned}
\end{equation}

For Eq. \eqref{eq:a1bound} we can see that $a_1<0$ under
\textbf{Condition 2a:}
\begin{align*}
\frac{\lambda_S}{\lambda_D}&>\frac{(1+\delta)N-1}{(1-\delta)N-1}.&\Box
\end{align*}

\textbf{Lemma 1$'$}

$a_0$ is negative under Condition 2a.

\textbf{Proof:}

\begin{equation}\label{eq:a0bound}
    \begin{aligned}
        a_0 &=
    \underline{P_0^S}\overline{P_2^D}-\overline{P_0^D}\underline{P_2^S}\\
    &=
    (1-\lambda_S)^{(1+\delta)N}\frac{(1+\delta)N[(1+\delta)N-1]}{2}\lambda_D^2(1-\lambda_D)^{(1+\delta)N-2}\\
    &- (1-\lambda_D)^{(1-\delta)N}\frac{(1-\delta)N[(1-\delta)N-1]}{2}\lambda_S^2(1-\lambda_S)^{(1-\delta)N-2}\\
    &=\frac{1}{2}(1-\lambda_S)^{(1-\delta)N-2}(1-\lambda_D)^{(1-\delta)N}\{(1-\lambda_S)^{2\delta N+2}(1+\delta)N[(1+\delta)N-1]\\
    &\cdot \lambda_D^2(1-\lambda_D)^{2\delta
    N-2}-(1-\delta)N[(1-\delta)N-1]\lambda_S^2\}\\
    &<\frac{1}{2}(1-\lambda_S)^{(1-\delta)N-2}(1-\lambda_D)^{(1-\delta)N}\{(1+\delta)N[(1+\delta)N-1]\lambda_D^2\\
    &-(1-\delta)N[(1-\delta)N-1]\lambda_S^2\}\\
    &=\frac{1}{2}(1-\lambda_S)^{(1-\delta)N-2}(1-\lambda_D)^{(1-\delta)N}\{[(1+\delta)N-1]^2\lambda_D^2+[(1+\delta)N-1]\lambda_D^2\\
    &- [(1-\delta)N-1]^2\lambda_S^2-[(1-\delta)N-1]\lambda_S^2\}\\
    &<0
    \end{aligned}
\end{equation}
In the last step, we made use of Condition 2a. $\Box$

\textbf{Lemma 2}

$a_2(n)$ is positive under \textbf{Condition 2:}
$$
\frac{\lambda_S}{\lambda_D}>\frac{(1+\delta)N-2}{(1-\delta)N-2}\left[\frac{(1+\delta)N-2}{2\delta
N}\right]^\frac{2\delta
N}{(1-\delta)N-2}\left[\frac{(1+\delta)N-2}{(1-\delta)N-2}\cdot\frac{e^2}{2\delta
N}\right]^\frac{1}{2[(1-\delta)N-2]}.
$$

\textbf{Proof:}

Expanding Eq. \eqref{eq:a2}, note that $3\le n \le (1-\delta)N$, we
have
\begin{equation}\label{eq:a2bound}
    \begin{aligned}
    a_2(n) &= \underline{P_n^S}\overline{P_2^D}-\overline{P_n^D}\underline{P_2^S}\\
        &= {(1-\delta)N \choose
        n}\lambda_S^n(1-\lambda_S)^{(1-\delta)N-n}\frac{(1+\delta)N[(1+\delta)N-1]}{2}\lambda_D^2(1-\lambda_D)^{(1+\delta)N-2}\\
        &- {(1+\delta)N \choose
        n}\lambda_D^n(1-\lambda_D)^{(1+\delta)N-n}\frac{(1-\delta)N[(1-\delta)N-1]}{2}\lambda_S^2(1-\lambda_S)^{(1-\delta)N-2}\\
        &=
        \lambda_S^2\lambda_D^2(1-\lambda_S)^{(1-\delta)N-n}(1-\lambda_D)^{(1+\delta)N-n}\frac{[(1-\delta)N]![(1+\delta)N]!}{2\cdot
        n!}[b_1(n)-b_2(n)],
    \end{aligned}
\end{equation}
where
\begin{align*}
    b_1(n) &=
    \frac{\lambda_S^{n-2}(1-\lambda_D)^{n-2}}{[(1-\delta)N-n]![(1+\delta)N-2]!}>0,\\
    b_2(n) &=
    \frac{\lambda_D^{n-2}(1-\lambda_S)^{n-2}}{[(1+\delta)N-n]![(1-\delta)N-2]!}>0.
\end{align*}
To show that $a_2(n)>0$, we need to show that $b_1(n)>b_2(n)$. Since
they are both positive, we could try to show that $b_1(n)/b_2(n)>1$.
\begin{align*}
    \frac{b_1(n)}{b_2(n)} &=
    \frac{[(1+\delta)N-n]![(1-\delta)N-2]!}{[(1+\delta)N-2]![(1-\delta)N-n]!}\left[\frac{\lambda_S(1-\lambda_D)}{\lambda_D(1-\lambda_S)}\right]^{n-2}\\
    &=
    \prod_{i=3}^n\left[\frac{(1-\delta)N-i+1}{(1+\delta)N-i+1}\cdot\frac{\lambda_S(1-\lambda_D)}{\lambda_D(1-\lambda_S)}\right].
\end{align*}
Define the last term of the product as
$$
d(n) =
\frac{(1-\delta)N-n+1}{(1+\delta)N-n+1}\cdot\frac{\lambda_S(1-\lambda_D)}{\lambda_D(1-\lambda_S)},
$$
which is a decreasing function of $n$. Note that $d(n)$ is always
positive. Due to the decreasing nature of $d_n$ on $n$, there exists
a real number $n_0$ satisfying the following criterium: for any
$n<n_0$, $d(n)>1$; for any $n\ge n_0$, $d(n)\ge1$. We can easily see
the following facts:

1) If $n<n_0$, we know for sure that $b_1(n)/b_2(n)>1$, which means
$a_2(n)>0$.

2) If $n\ge n_0$, $b_1(n)/b_2(n)$ decreases as $n$ increases. Since
$n\le(1-\delta)N$, we have
\begin{align*}
    \frac{b_1(n)}{b_2(n)} &=
    \prod_{i=3}^n\left[\frac{(1-\delta)N-i+1}{(1+\delta)N-i+1}\cdot\frac{\lambda_S(1-\lambda_D)}{\lambda_D(1-\lambda_S)}\right]\\
    &\ge
    \prod_{i=3}^{(1-\delta)N}\left[\frac{(1-\delta)N-i+1}{(1+\delta)N-i+1}\cdot\frac{\lambda_S(1-\lambda_D)}{\lambda_D(1-\lambda_S)}\right]\\
    &= \frac{[(1-\delta)N-2]!(2\delta
    N)!}{[(1+\delta)N-2]!}\left[\frac{\lambda_S(1-\lambda_D)}{\lambda_D(1-\lambda_S)}\right]^{(1-\delta)N-2}\\
    &= \frac{1}{{(1+\delta)N-2 \choose 2\delta N}}\left[\frac{\lambda_S(1-\lambda_D)}{\lambda_D(1-\lambda_S)}\right]^{(1-\delta)N-2}.
\end{align*}

Therefore $a_2(n)>0$ under \textbf{Condition 2b:}
$$
\frac{\lambda_S}{\lambda_D}>{(1+\delta)N-2 \choose 2\delta
N}^{\frac{1}{(1-\delta)N-2}}.
$$
Note that $N$ is usually very large, which means the evaluation of
 Condition 2b can be computationally challenging. To simplify this condition,
 we can make use of
Stirling's approximation
$$
\sqrt{2\pi}n^{n+\frac{1}{2}}\exp(-n+\frac{1}{12n+1})<n!<\sqrt{2\pi}n^{n+\frac{1}{2}}\exp(-n+\frac{1}{12n}),
$$
which can be simplified to be
\begin{equation}\label{eq:stirling}
n^{n+\frac{1}{2}}e^{-n}<n!<n^{n+\frac{1}{2}}e^{-n+1}.
\end{equation}
With the help of Eq. \eqref{eq:stirling}, we can derive a simpler
and stronger version of Condition 2b:

\textbf{Condition 2:}
$$
\frac{\lambda_S}{\lambda_D}>\frac{(1+\delta)N-2}{(1-\delta)N-2}\left[\frac{(1+\delta)N-2}{2\delta
N}\right]^\frac{2\delta
N}{(1-\delta)N-2}\left[\frac{(1+\delta)N-2}{(1-\delta)N-2}\cdot\frac{e^2}{2\delta
N}\right]^\frac{1}{2[(1-\delta)N-2]}.\Box
$$

Note that Condition 2 is also stronger than Condition 2a. Therefore
Lemma 1 is also true under Condition 2.

\textbf{Lemma 2$'$}

$\sum_{n=3}^{(1-\delta)N}a_2(n)Z_2(n)\ge0$ under Condition 2.

\textbf{Proof:}

From Eq. \eqref{eq:Z2} we can clearly see that $Z_2(n)\ge0$. $\Box$

\textbf{Lemma 3}

$$a_3> -\frac{2\delta
   N(1-\lambda_D)^{2\delta
   N-1}\underline{P_2^S}}{[(1-\delta)N+1]!}.$$

\textbf{Proof:}

Expand Eq. \eqref{eq:a3}, we have
\begin{equation}\label{eq:a3bound}
\begin{aligned}
   a_3 &=
   -\sum_{n=(1-\delta)N+1}^{(1+\delta)N}\overline{P_n^D}\underline{P_2^S}Z_3(n)\\
   &\ge
   -\sum_{n=(1-\delta)N+1}^{(1+\delta)N}\overline{P_n^D}\underline{P_2^S}&(\because 0\le Z_3(n)\le 1)\\
   &\ge -2\delta N\overline{P_{(1-\delta)N+1}^D}\underline{P_2^S}&(\because 0\le \overline{P_n^D}<\overline{P_{(1-\delta)N+1}^D})\\
   &= -2\delta N {(1+\delta)N \choose
   (1-\delta)N+1}\lambda_D^{(1-\delta)N+1}(1-\lambda_D)^{2\delta
   N-1}\underline{P_2^S}\\
   &= -2\delta
   N\frac{1}{[(1-\delta)N+1]!}(1-\lambda_D)^{2\delta
   N-1}\underline{P_2^S}\prod_{i=0}^{(1-\delta)N}\{[(1+\delta)N-i]\lambda_D\}\\
   &> -\frac{2\delta
   N(1-\lambda_D)^{2\delta
   N-1}\underline{P_2^S}}{[(1-\delta)N+1]!}.&(\because [(1+\delta)N-i]\lambda_D<1)\\
\Box
\end{aligned}
\end{equation}

Note that $|a_3|$ is in the order of $O(\frac{1}{N!})$. It is very
close to 0.

From Eqs. \eqref{eq:qs-qd}-\eqref{eq:a3bound} we can conclude that
\begin{equation*}
    \begin{aligned}
    Z_1 &\ge
    \frac{Q^D\underline{P_2^S}-Q^S\overline{P_2^D}+a_0\overline{Q^V}+\sum_{n=3}^{(1-\delta)N}a_2(n)Z_2(n)+a_3}{-a_1}\\
    &> \frac{\underline{Q^D}\underline{P_2^S}-\overline{Q^S}\overline{P_2^D}+a_0\overline{Q^V}-\frac{2\delta
   N(1-\lambda_D)^{2\delta
   N-1}\underline{P_2^S}}{[(1-\delta)N+1]!}}{-a_1}=\underline{Z_1}.
    \end{aligned}
\end{equation*}
under \textbf{Condition 2:}
$$
\frac{\lambda_S}{\lambda_D}>\frac{(1+\delta)N-2}{(1-\delta)N-2}\left[\frac{(1+\delta)N-2}{2\delta
N}\right]^\frac{2\delta
N}{(1-\delta)N-2}\left[\frac{(1+\delta)N-2}{(1-\delta)N-2}\cdot\frac{e^2}{2\delta
N}\right]^\frac{1}{2[(1-\delta)N-2]}.
$$

Therefore the lower bound of $Q_1^S$ is given by
\begin{align*}
Q_1^S &\ge \underline{P_1^S}Z_1>\underline{P_1^S}\underline{Z_1}\\
&=
\underline{P_1^S}\frac{\underline{Q^D}\underline{P_2^S}-\overline{Q^S}\overline{P_2^D}+(\underline{P_0^S}\overline{P_2^D}-\overline{P_0^D}\underline{P_2^S})\overline{Q^V}-\frac{2\delta
   N(1-\lambda_D)^{2\delta
   N-1}\underline{P_2^S}}{[(1-\delta)N+1]!}}{\overline{P_1^D}\underline{P_2^S}-\underline{P_1^S}\overline{P_2^D}} = \underline{Q_1^S}
\end{align*}

This completes our proof of Proposition 1.

\section{Estimate of $e_1^S$}\label{app:e1}
The derivation of the upper bound of $e_1^S$ is relatively simpler
than that of the lower bound of $Q_1^S$. Similar as Eq.
\eqref{eq:qs-qd} we have
$$
E^S\cdot
Q^S=\sum_{m=(1-\delta)N}^{(1+\delta)N}P_\text{in}(m)\sum_{n=0}^\infty
P_n^S(m)Y_{m,n}e_{m,n},
$$
where $e_{m,n}$ is the error rate for signals with $m$ input photons
and $n$ output photons. Rearranging terms, we have

\begin{equation}\label{eq:q1se1s}
\begin{aligned}
    Q_1^Se_1^S &= E^S\cdot Q^S - Q_0^Se_0^S - \sum_{n=2}^\infty
    Q_n^Se_n^S\\
                &\le E^S\cdot Q^S - Q_0^Se_0^S\\
                &= E^S\cdot Q^S -
                \sum_{m=(1-\delta)N}^{(1+\delta)N}P_\text{in}(m)P_0^S(m)Y_{m,0}e_{m,0}\\
                &\le E^S\cdot Q^S -
                \underline{P_0^S}\sum_{m=(1-\delta)N}^{(1+\delta)N}P_\text{in}(m)Y_{m,0}e_{m,0}\\
                &= E^S\cdot Q^S - \underline{P_0^S}E^V\cdot Q^V.
\end{aligned}
\end{equation}
The upper bound of $e_1^S$ is thus given by
$$
    e_1^S\le \frac{E^S\cdot Q^S - \underline{P_0^S}E^V\cdot Q^V}{Q_1^S}\le\frac{\overline{E^S\cdot Q^S} -
    \underline{P_0^S}\underline{E^V \cdot Q^V}}{\underline{Q_1^S}}.
$$

This completes our proof of Proposition 2.

\section{The one-decoy protocol}\label{app:one-decoy}
In one-decoy protocol, there is no vacuum state. Therefore we cannot
measure $Q_e^V$ or $E_e^V$. If we still want to estimate
$\underline{Q_1^S}$ via Eq. \eqref{eq:q1boundWV} and
$\overline{e_1^S}$ via Eq. \eqref{eq:e1boundWV}, we need to estimate
$\overline{Q_V}$ in Eq. \eqref{eq:q1boundWV} and $\underline{E^V
\cdot Q^V}$ in Eq. \eqref{eq:e1boundWV} in another way.

To estimate $\overline{Q^V}$, we can look into Eq.
\eqref{eq:q1se1s}:
$$
\underline{P_0^S}E^VQ^V\le E^SQ^S-Q_1^Se_1^S\le
E^SQ^S\le\overline{E^SQ^S}.
$$
Therefore
\begin{equation}\label{eq:qv_onedecoy}
    Q^V\le\frac{\overline{E^SQ^S}}{\underline{P_0^S}E^V}=\overline{Q^V},
\end{equation}
where $\overline{E^SQ^S}$ can be estimated from Eqs.
\eqref{eq:Ebound}, $\underline{P_0^S}$ can be estimated from Eqs.
\eqref{eq:pnbound}, and $E^V=0.5$ in asymptotic case.

Plugging Eq. \eqref{eq:qv_onedecoy} into Eq. \eqref{eq:q1boundWV},
we have the expression of $Q_1^S$ with the one-decoy protocol:
\begin{equation*}
    Q_1^S > \underline{Q_1^S}= \underline{P_1^S}\frac{\underline{Q^D}\underline{P_2^S}-\overline{Q^S}\overline{P_2^D}+(\underline{P_0^S}\overline{P_2^D}-\overline{P_0^D}\underline{P_2^S})\frac{\overline{E^SQ^S}}{\underline{P_0^S}E^V}-\frac{2\delta
   N(1-\lambda_D)^{2\delta
   N-1}\underline{P_2^S}}{[(1-\delta)N+1]!}}{\overline{P_1^D}\underline{P_2^S}-\underline{P_1^S}\overline{P_2^D}}
\end{equation*}

As for the estimate of $\underline{E^V \cdot Q^V}$, we can simply
use the following fact: $E^V \cdot Q^V \ge 0$. Therefore the
expression of $\overline{e_1^S}$ in one-decoy protocol is given by

$$
    e_1^S\le\overline{e_1^S} =\frac{\overline{E^S\cdot Q^S}}{\underline{Q_1^S}}.
$$

This completes our proof of Proposition 3.

\end{appendix}

\end{document}